\RequirePackage[2020/10/01]{latexrelease}
\documentclass[aip,reprint]{revtex4-1}

\usepackage{graphicx}
\usepackage{dcolumn}
\usepackage{bm}
\usepackage[dvipsnames]{xcolor}
\usepackage{hyperref}
\usepackage{amsmath}
\usepackage[normalem]{ulem}

\begin{document}

\title{Parametric study of the polarization dependence of nonlinear Breit-Wheeler pair creation process using two laser pulses}

\author{Qian Qian}
\email{qqbruce@umich.edu}
\affiliation{G\'{e}rard Mourou Center for Ultrafast Optical Science, University of Michigan, 2200 Bonisteel Boulevard, Ann Arbor, Michigan 48109, USA}

\author{Daniel Seipt}
\affiliation{Helmholtz Institute Jena, Fr$\ddot{o}$belstieg 3, 07743 Jena, Germany}

\author{Marija Vranic}
\affiliation{GoLP/Instituto de Plasmas e Fusão Nuclear, Instituto Superior Técnico, Universidade de Lisboa, 1049-001 Lisbon, Portugal}

\author{Thomas E. Grismayer}
\affiliation{GoLP/Instituto de Plasmas e Fusão Nuclear, Instituto Superior Técnico, Universidade de Lisboa, 1049-001 Lisbon, Portugal}

\author{Thomas G. Blackburn}
\affiliation{Department of Physics, University of Gothenburg, SE-41296 Gothenburg, Sweden} 

\author{Christopher P. Ridgers}
\affiliation{York Plasma Institute, Department of Physics, University of York, York, YO10 5DD, United Kingdom}

\author{Alexander G. R. Thomas}
\affiliation{G\'{e}rard Mourou Center for Ultrafast Optical Science, University of Michigan, 2200 Bonisteel Boulevard, Ann Arbor, Michigan 48109, USA}

\date{\today}

\begin{abstract}
With the rapid development of high-power petawatt class lasers worldwide, exploring physics in the strong field QED regime will become one of the frontiers for laser-plasma interactions research. Particle-in-cell codes, including quantum emission processes, are powerful tools for predicting and analyzing future experiments where the physics of relativistic plasma is strongly affected by strong-field QED processes. The spin/polarization dependence of these quantum processes has been of recent interest. In this article, we perform a parametric study of the interaction of two laser pulses with an ultrarelativistic electron beam. The first pulse is optimized to generate high-energy photons by nonlinear Compton scattering and efficiently decelerate electron beam through quantum radiation reaction. The second pulse is optimized to generate electron-positron pairs by nonlinear Breit-Wheeler decay of photons with the maximum polarization dependence. This may be experimentally realized as a verification of the strong field QED framework, including the spin/polarization rates.   
\end{abstract}

\pacs{}

\maketitle 

\section{Introduction}
Strong field quantum electrodynamics (SF QED) processes occur when electromagnetic fields exceed the quantum critical field strength \cite{Ritus_1985}.  Many high-power laser facilities constructed  worldwide over the last decade now aim to reach the regime where strong field QED effects can appear \cite{Danson_Laser_Sci_Eng}. Even though  current laser technology is not able to reach the QED critical field strength  in the laboratory, because strong field QED effects depend on Lorentz invariant parameters, including the electric field in the rest frame of the particles \cite{Burke_PRL_1997}, such a regime is accessible with lower intensity lasers when relativistic plasma flows interact with the fields. 
QED theory has been well verified at the single particle level, but the physics  becomes more complicated when  the QED processes are coupled with relativistic plasma dynamics. These can lead to phenomena such as bright gamma-ray emission\cite{Nakamura_PRL_2012_gamma_ray, Ridgers_PRL_2012_gamma_ray, Hirotani_APJ_2016_astro_gamma_ray, Frank_PLB_1989_astro_gamma_ray, Liangqi_POP_2021} and quantum radiation reaction \cite{Cole_PRX_2018_QRR, Poder_PRW_2018_QRR,
Piazza_PRL_2009_QRR, Piazza_PRL_2010_QRR, Piazza_RMP_2012_QRR, 
Gonoskov_RMP_2022_QRR,
Fedotov_PR_2023_QRR,
Thomas_PRX_2012_QRR, Zhang_NJP_2015_QRR, Blackburn_PRL_2018_QRR, Vranic_NJP_2016, Niel_PRE_2018, Ridgers_JPP_2017}, electron-positron pair showers \cite{Tsai_PRD_1993_shower, Mironov_PLA_2014_two_types_cascade, Bulanov_PRA_2013_shower, Sokolov_PRL_2010_shower, Mercuri-Baron_NJP_2021}, and avalanches \cite{Mironov_PLA_2014_two_types_cascade, Fedotov_PRL_2010_avalanche,  Bell_PRL_2008, Seipt_NJP_2021, Kirk_PPCF_2009_avalanche, Elkina_PRSTAB_2011_avalanche, Grismayer_POP_2016_avalanche, Grismayer_PRE_2017_avalanche, Vranic_PPCF_2017_avalanche, Song_NJP_2021, Luo_SR_2018, Luo_POP_2018, Luo_POP_2015, Tang_PRA_2014}.
As a result, particle-in-cell (PIC) codes modified to include the QED processes become a useful tool for exploring the plasma dynamics and SF QED effects in supercritical fields. Many state-of-the-art PIC  codes have included the QED module \cite{Ridgers_JCP, PICADOR_QED_PIC, Grismayer_PRE_2017_avalanche} using the Local Constant Field Approximation (LCFA), under which the quantum processes reduce to probabilistic emissions of either  with a rate that only depends on a parameter $\chi_q$, which corresponds to the rest-frame electric field experienced by a relativistic lepton divided by the critical field. 

The standard QED-PIC algorithm averages over spin and polarization rates in their calculations. The strong field QED processes are, however, fundamentally spin and polarization-dependent \cite{Sokolov_Synchrotron, Ivanov_QED_spin, Ivanov_QED_spin2, Sorbo_PRA, Seipt_PRA_2018, Sorbo_PPCF, Seipt_PRA_2020, Seipt_NJP_2021, Chen_PRL_2019, Li_PRL_2019, Li_PRL_2020_gamma_ray, Li_PRL_2020_helicity_transfer, Wan_PRR_2020, Guo_PRR_2020, Dai_MRE_2022, Chen_PRD_2022, Blackburn_arXiv_2023}. Spin and polarization are important quantities in particle physics research. For high-energy lepton colliders, collisions between spin-polarized electron and positron beams are preferred to study  possible new physics beyond the standard model \cite{Barish_JMPA, Pick_PR}. Recent studies relevant to high-power laser facilities have shown that the spin/polarization distinguished QED code can more accurately simulate multi-staged processes like the avalanche and shower-type electron-positron pair production cascade processes \cite{Seipt_PRA_2020, Seipt_NJP_2021, King_PRA_2013}, condition lepton beam spin distributions  \cite{Seipt_PRA_2019, Li_PRL_2019, Sorbo_PPCF}, and to be presented in plasma instabilities \cite{Gong_PRL_2023}. In Ref. \onlinecite{Wan_PRR_2020}, a method using two laser pulses was proposed;  polarized gamma rays are generated with a first pulse, and subsequently, a second pulse generates electron-positron pairs, with the yield varying with the relative polarization of the lasers because of the polarization dependent rates.
 
In this article, we outline a QED module implemented in the particle-in-cell code framework OSIRIS 4.0 \cite{Fonseca_Osiris_note, Fonseca_Osiris_PPCF, Vranic_CPC_2015}, which is modified to include spin and polarization dependent rates and use it to study the parameter regime of the two-pulse $\gamma$-polarimetry configuration proposed in [\onlinecite{Wan_PRR_2020}],
in particular, optimizing the yield and finding the parametric dependence on the laser pulse characteristics.  
We will briefly discuss the theoretical framework of the code and how the code is implemented. The algorithm was validated in constant field configurations, and a more rigorous benchmark to reproduce the main results of a seeded electron-positron pair cascade \cite{Bell_PRL_2008} in a rotating electric field calculated using a Boltzmann type solver\cite{Seipt_NJP_2021}. 
Finally, we present the two-pulse pair production scheme, which can be achieved in an all-optical experiment using laser wakefield accelerated electrons \cite{2020_roadmap_on_plasma_accelerators}. This scheme highlights the polarization dependency of the NBW pair production process.  We find the  parameters required to maximize the measurable difference in pair production yield when we rotate the laser polarization of the second pulse. By optimizing the process for the two pulses in terms of their pulse length and field strength, simulations using our spin and polarization-resolved PIC code demonstrated over 50\% difference in pair production yield by simply changing the polarization directions of two linearly polarized laser pulses. We also discuss the criteria for laser and electron beam parameters for designing an experiment based on this scheme.

\section{Background}
\subsection{Strong field QED processes and Plasma physics}
Strong field QED involves the physics of particles experiencing strong EM fields of order the QED critical field strength $E_{cr}$. This is the characteristic field strength that does $m_ec^2$ work over a (reduced) Compton wavelength: $eE_{cr}\lambdabar_c = m_ec^2$, or $E_{cr} = 1.32\times10^{18}$~Vm$^{-1}$.  One  important  measure of the importance of the strong field QED effects during an interaction is the quantum strength parameter for particle $q$, defined as $\chi_q = {\lvert\lvert F_{\mu\nu}p^{\nu}_{q} \rvert\rvert }/{(m_ecE_{cr})} \equiv  ||\gamma(\mathbf{E}+\mathbf{v}\times \mathbf{B})||/E_{cr}$. 
When $\chi_q\gg 1$,  strong field QED effects will 
be substantial. From the definition, we could find that the quantum parameter \emph{for leptons} is related to the ratio of the electric field in its rest frame. For photons, there is no straightforward interpretation. 
As a result, for a lepton moving close to the speed of light with relativistic parameter $\gamma\gg 1$, the field strength it sees in its rest frame will be boosted by a factor of $\sim\gamma$, and so the quantum parameter will be large, $\chi_q\gg1$, even for laboratory fields that are weak compared to $E_{cr}$. 

Our study focuses on the two lowest-order processes in strong field QED: Nonlinear Compton scattering (NLC)\cite{Nikishov_JETP_1964, Brown_PRL_1964, Narozhny_JETP_1965, Bula_PRL_1996}, and Nonlinear Breit Wheeler process (NBW)\cite{Narozhny_JETP_1965, Reiss_JMP_1962, Burke_PRL_1997}. 
Compared with high-order quantum processes, the probabilities of these lowest-order processes are dominant by a factor of order $\alpha=1/137$. In most cases, we could ignore the influence of the high-order processes. Furthermore, we work with local constant field approximation (LCFA) \cite{Ridgers_JCP} rates, which requires a weak field approximation, i.e., that the laboratory fields are weak compared to $E_{cr}$. This condition means, first, that the fields in the rest frame  approximate crossed fields, and so the rates need only to be calculated for a crossed field configuration. Second, because the fields are weak and therefore, for quantum processes of interest, the initial  and final particles need to be energetic. The ``formation length'' of such processes \cite{Baier_Katkov} is short compared to the scales of  spatiotemporal fluctuations in the fields in practice, so they can be considered to be constant, and the emitted particles can be considered to be collinear with the initial particle momentum. Hence, the rates only need calculating for constant and crossed electromagnetic field configurations. For the interaction with a laser field, this approximation is effectively valid if the normalized field strength $a_0 = eE_0/m_ec\omega\gg 1$, where $E_0$ is the peak electric field of the laser, while the field is much smaller than the critical field strength, $E/E_{cr}\ll1$. This implies that two normalized Lorentz invariants during the interaction: $\mathcal{S} = \left(E^2-c^2B^2\right)/E_{cr}^2$ and  $\mathcal{P} = |{\pmb E} \cdot c{\pmb B}|/E_{cr}^2$ are much smaller than 1 \cite{Dinu_PRL, Piazza_PRA}. These approximations can break down \cite{Blackburn_LCFA_breakdown, Ilderton_LCFA_breakdown}, so it is important to consider when this numerical LCFA framework is applicable. For example, at the transition region where $a_0 \sim 1$, the LCFA approximation will not be valid. We need to use other approximations like the locally monochromatic approximation (LMA)\cite{Bamber_PRD_1999, Chen_1995_CAIN, Hartin_IntJMPA_2018, Heinzl_PRA_2020, Torgrimsson_NJP_2021}, which also recently looked into the polarization-dependent QED effect\cite{Blackburn_arXiv_2023}.  

Strong field QED effects can become important in relativistic plasma dynamics. In particular, the NLC process  affects the dynamics of the charged particles through radiation back-reaction when emitting an energetic photon. The NBW process modifies the plasma density by absorbing  high-energy photons and generating electron-positron pairs. When the fields are strong enough, and the relativistic plasma is energetic enough to keep these quantum effects continuously changing the basic plasma parameters, they could finally influence the plasma's collective behavior. The plasma dynamics can be different from the classical situation.
This coupling system is defined in some literature as a `QED-plasma' \cite{Zhang_POP, melrose_2008_quantum, melrose_2013_quantum, Uzdensky_2014_RPP, Uzdensky_2019_arXiv}. 
It naturally appears in extreme astrophysical environments, including neutron star atmospheres \cite{Goldreich_Astrophys, Cruz_AJ_2021_Neutron}, pulsar magnetospheres \cite{Cruz_AJ_2021_Pulsar} and black hole environments \cite{Ruffini_Astrophys}. To produce this QED plasma in the lab is one of the ambitions for high-energy laser facilities. The possible rich phenomena inside this system need to be better understood. 
\subsection{Spin and polarization-dependent QED}
\label{Sec_background_spin_qed}
\begin{figure}[h]
\includegraphics{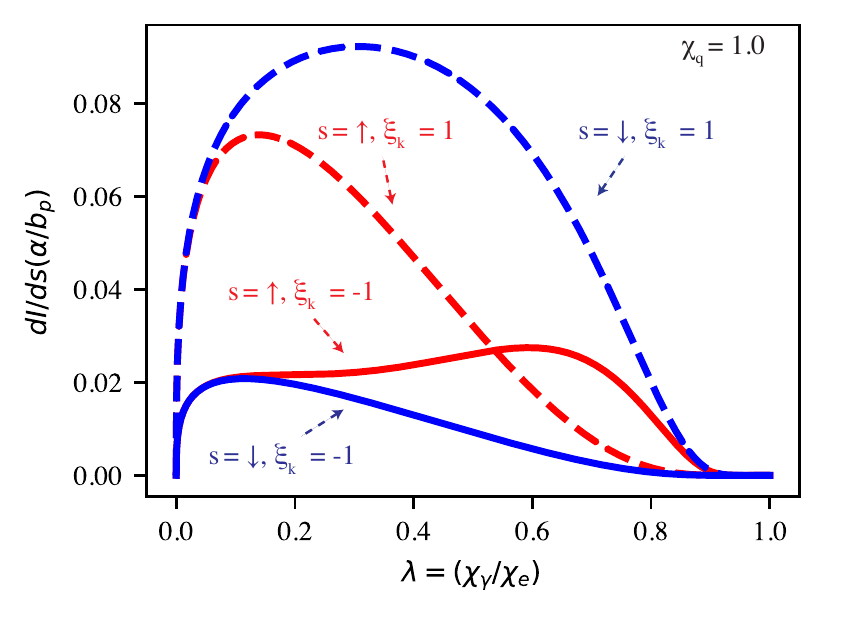}
\centering
\caption{Spin and polarization involved-NLC spectra for lepton with initial spin state $s$. $\xi_k$ is the Stokes parameter of the radiated photon.}
\label{fig:NLC_spect}
\end{figure}

\begin{figure}[h]
\includegraphics{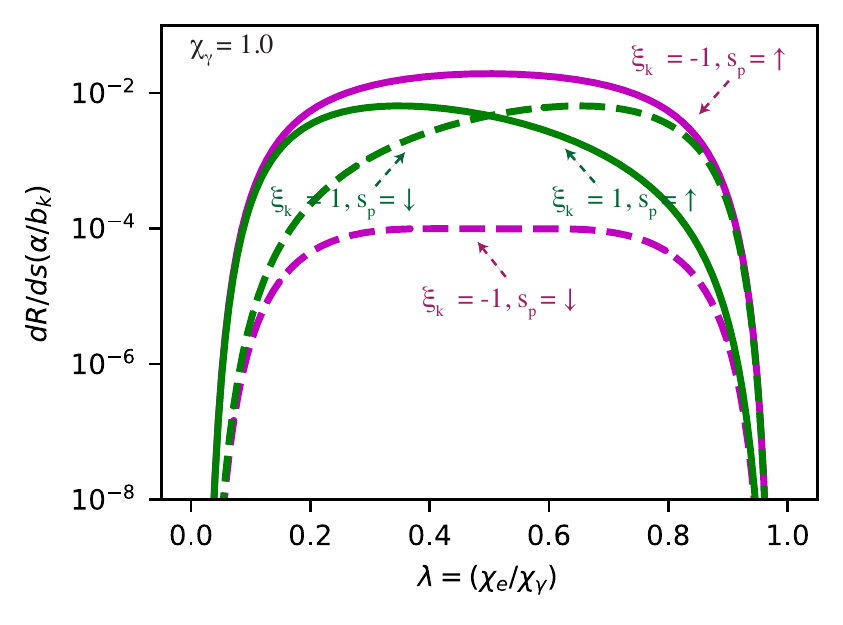}
\centering
\caption{Spin and polarization involved-NBW spectrums for the photon with Stokes parameter $\xi_k$. The generated positron spin state is $s_p$.}
\label{fig:NBW_spect}
\end{figure}

Most of the studies for QED plasma use a PIC code with spin averaged QED rates \cite{Ridgers_JCP, PICADOR_QED_PIC, Grismayer_PRE_2017_avalanche}, in which the QED processes only depend on the momentum of the particles and the EM field they experience. Fundamentally, the quantum emission and pair production processes are all spin-dependent. The spin of the particles evolves both through precession in the fields and due to radiative spin transitions. We use a representation of the spin dynamics where a  vector representing the three components of the classical spin-polarization vector, representing the expectation over many measurements, evolves via the Thomas-Bargmann-Michel-Telegdi (T-BMT) equations, and radiative spin transitions are represented by a Monte-Carlo sampling algorithm where point-like photon emissions result in the spin vector collapsing onto an eigenstate of the local non-precessing (rest frame magnetic field) direction. This model is an incomplete description of the spin dynamics, in general\cite{Chen_PRD_2022},  but is exact when the leptons are initially unpolarized and in fixed direction fields like a linearly polarized laser field. Fig.~\ref{fig:NLC_spect} shows the radiation spectrum of the Nonlinear Compton scattering (NLC) process for an initial lepton spin state  $s =\ \uparrow$ or $\downarrow$
and the radiated photon Stokes parameter is $\xi_k = \pm 1$ when the quantum parameter is $\chi_q = 1.0$. The up or down arrow indicates that the spin is either parallel or antiparallel to the rest frame magnetic field in this case. The Stokes parameter here will be explained in detail in Section \ref{spin_basis}. We can see that a lepton's initial spin state will modify the radiation spectrum and the radiated photon's polarization state. Fig.~\ref{fig:NBW_spect} shows the generated positron energy spectrum from the Nonlinear Breit Wheeler (NBW) process for a photon with Stokes parameter $\xi_k = \pm 1$ and the generated positron spin state $s_{p} =\ \uparrow$ or $\downarrow$. Again, the photon's polarization state will influence the generated lepton's energy spectrum and spin state. 
Note that the spectra are asymmetrical for different photon polarization and lepton spin states \cite{Seipt_PRA_2020, Seipt_NJP_2021}. This asymmetry in the spectra allows us to explore possible regimes for generating polarized gamma-ray and lepton bunches using  strong field QED process \cite{Sorbo_PPCF, Sorbo_PRA, Li_PRL_2019}.



\subsection{Lepton spin and photon polarization basis }
\label{spin_basis}
The strong field QED model we discussed so far follows the local constant field approximation (LCFA). For  strong field QED processes $\chi_q\gg1$, the electric and magnetic fields in the rest frame of a highly relativistic particle will be of order $E_{cr}$, while the LCFA requires $\mathcal{P}\ll 1$. As a result, the direction of the electric and magnetic fields in the particle's rest frame and the momentum vector should be close to mutually perpendicular. 
We may use this inherently orthogonal system to construct a spin and polarization basis. Here, because the PIC code uses three-vector objects, we express these in  three-vector  instead of  four-vector form. There are three mutually orthogonal vectors, here: ($\boldsymbol\varepsilon$, ${\pmb{\beta}}$, $\pmb{\kappa}$), where  ${\pmb{\varepsilon}} = {{\pmb{E}_{RF}}}/|{\pmb{E}_{RF}|}$ is a unit vector in the direction of the rest frame electric field, $\pmb{\beta} = {\pmb{B}_{RF}}/{|\pmb{B}_{RF}|}$ is a unit vector in the direction of the rest frame magnetic field, and $\pmb{\kappa} = \pmb\varepsilon\times \pmb\beta$ is a unit vector that agrees with the direction of the background field Poynting vector.

In a general field configuration, the lepton spin and photon polarization components combining all three of these orthogonal directions need to be accounted for \cite{BKS, Chen_PRD_2022}. The emitted photon polarization can be in combinations of linear and circularly polarized modes. 
The approximation that the laser electric field is unidirectional starts to break down for an interaction with a tightly focused laser pulse, where the longitudinal field components become significant. However, for the configuration explored in this study involving a linearly polarized laser with moderate focal spot size ($w_0 \sim 10\lambda$) 
colliding with electrons / emitted photons and assuming the electrons are initially in an unpolarized state prior to interaction with the laser fields, the polarization of the emitted photons is restricted to linear polarized states \cite{Seipt_PRA_2020}. The electron spin only gains a component along the magnetic field direction in the emission process.
\subsubsection{Photon polarization basis}
Using the basis vectors $({\pmb\varepsilon}, \pmb{\beta}, \pmb{\kappa})$, we can define a (linear) polarization basis for a photon with its three-momentum along unit vector $\hat{\pmb{k}}$ with polarization vectors
\begin{equation*}
    \pmb\Lambda_1 = \pmb\varepsilon + \frac{\hat{\pmb{k}}\cdot \pmb\varepsilon}{1-\hat{\pmb{k}}\cdot \pmb\kappa}\left(\pmb\kappa-\hat{\pmb{k}}\right), \ \ \ \pmb\Lambda_2 = \pmb\beta + \frac{\hat{\pmb{k}}\cdot \pmb\beta}{1-\hat{\pmb{k}}\cdot \pmb\kappa}\left(\pmb\kappa-\hat{\pmb{k}}\right)\;.
\end{equation*}
Especially when $\hat{\pmb{k}} = \pm \pmb\kappa $, $\pmb\Lambda_1 = \pmb \varepsilon$, $\pmb\Lambda_2 = \pmb \beta$. By construction, the polarization basis vectors fulfill $\hat{\pmb{k}} \cdot \pmb\Lambda_i = 0$ and $\pmb\Lambda_i\cdot\pmb\Lambda_j = \delta_{ij}$. An arbitrarily polarized photon (in a pure state) with polarization vector $\pmb\epsilon_k$ can therefore be written as the superposition
\begin{equation*}
    \pmb\epsilon_k = c_1\pmb\Lambda_1 + c_2\pmb\Lambda_2
\end{equation*}
We characterize the photon polarization state using the Stokes parameter $\xi_k = |c_1|^2 - |c_2|^2$,
where, in general, $\xi_k\in [-1, 1]$. Here, we consider that photons to be produced in an eigenstate of the polarization operator. $\xi_k$, is chosen to be an integer, and $\xi_k \in \{-1, 1\}$.

\begin{figure}[h]
\includegraphics[width=6.5cm]{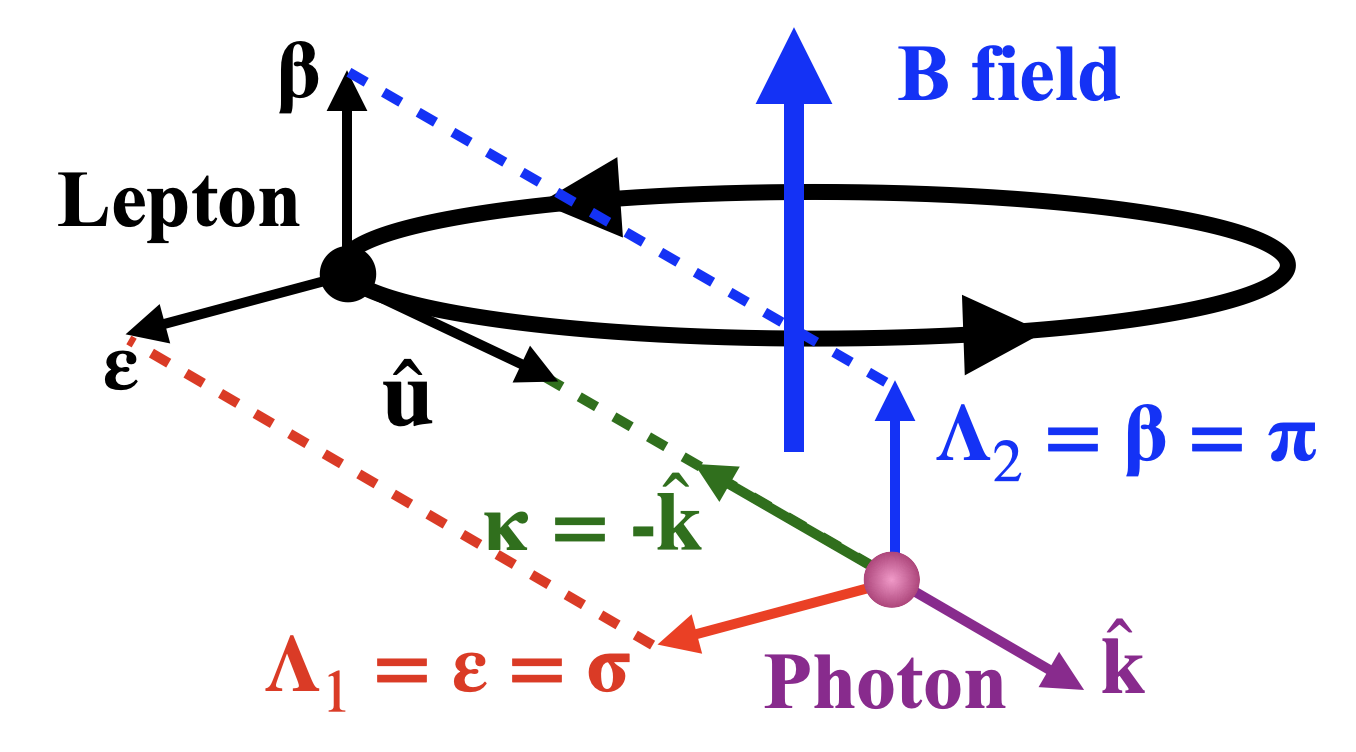}
\centering
\caption{Photon polarization basis. A high energy lepton with momentum in direction $\hat{\pmb u}$ emits a photon whose momentum direction $\hat{\pmb k} = \hat{\pmb u}$. The polarization basis vector for the photon is $\pmb \sigma = \pmb\Lambda_1$, $\pmb \pi = \pmb\Lambda_2$. When $\pmb\kappa = -\hat{\pmb k}$, $\pmb\sigma$ and $\pmb\pi$ are in the same direction as $\pmb\varepsilon$ and $\pmb\beta$.}
\label{fig:synchrotron}
\end{figure}

This choice of the polarization basis coincides with the synchrotron radiation geometry \cite{Sokolov_Synchrotron}(Fig.~\ref{fig:synchrotron}). An electron gyrating inside a constant magnetic field will undergo synchrotron radiation. The two polarization directions $\pmb\sigma$ and $\pmb\pi$ generally used in synchrotron radiation are the same as the direction of $\pmb\Lambda_1$ and $\pmb\Lambda_2$. As a result, we refer to the photon polarized along the $\pmb\Lambda_1$ direction, which occurs more frequently, as a $\sigma$-polarized photon. And we refer to a  photon polarized along the $\pmb\Lambda_2$ direction as a $\pi$-polarized photon. 

\subsubsection{Lepton spin basis} 
During a quantum event, choosing a lepton spin basis with a direction vector $\pmb\zeta$ that doesn't precess in the background field both simplifies the calculation and results in a universal rate that can be calculated in a probabilistic way. The particle states are projected onto $\pmb\zeta$ to give a spin quantum number for the particle, $S_\zeta$. This local non-precessing spin vector during a quantum event is along the rest-frame magnetic field of the lepton \cite{Seipt_PRA_2020}, i.e.,  $\pmb\zeta = \pmb\beta$ and we only consider this component of the spin. 
Although, in general, spin precession may lead to other components of the spin vector, notably longitudinal polarization, the radiative processes only increase the spin component in the $\pmb\beta$ direction and only cause the $\pmb\epsilon$ and $\pmb\kappa$ components to decay away. Hence, this simplified model is applicable for situations including the one considered here, where an unpolarized lepton beam interacts with a plane electromagnetic field in which the magnetic field in the rest frame of the particle does not rotate direction (although it can oscillate in the negative and positive direction). 
\section{Code implementation}

QED PIC is typically implemented by coupling the QED processes, such as gamma-ray photon emission by leptons and pair production by gamma-ray photons, through a Monte Carlo algorithm to the classical PIC
code \cite{Ridgers_JCP}. In our spin and polarization-dependent QED PIC version of OSIRIS, we consider the influence of spin and polarization on the rate and spectrum of the quantum process. 
We also make use of the T-BMT
equations-based  spin pusher in the  PIC loop \cite{Vieira_PRAB} to track the classical spin precession between the quantum process. This includes the anomalous magnetic moment, which arises from the loop-level contributions to the quantum transitions \cite{Seipt_POP_2023}. 

\begin{figure*}
\includegraphics[width=15cm]{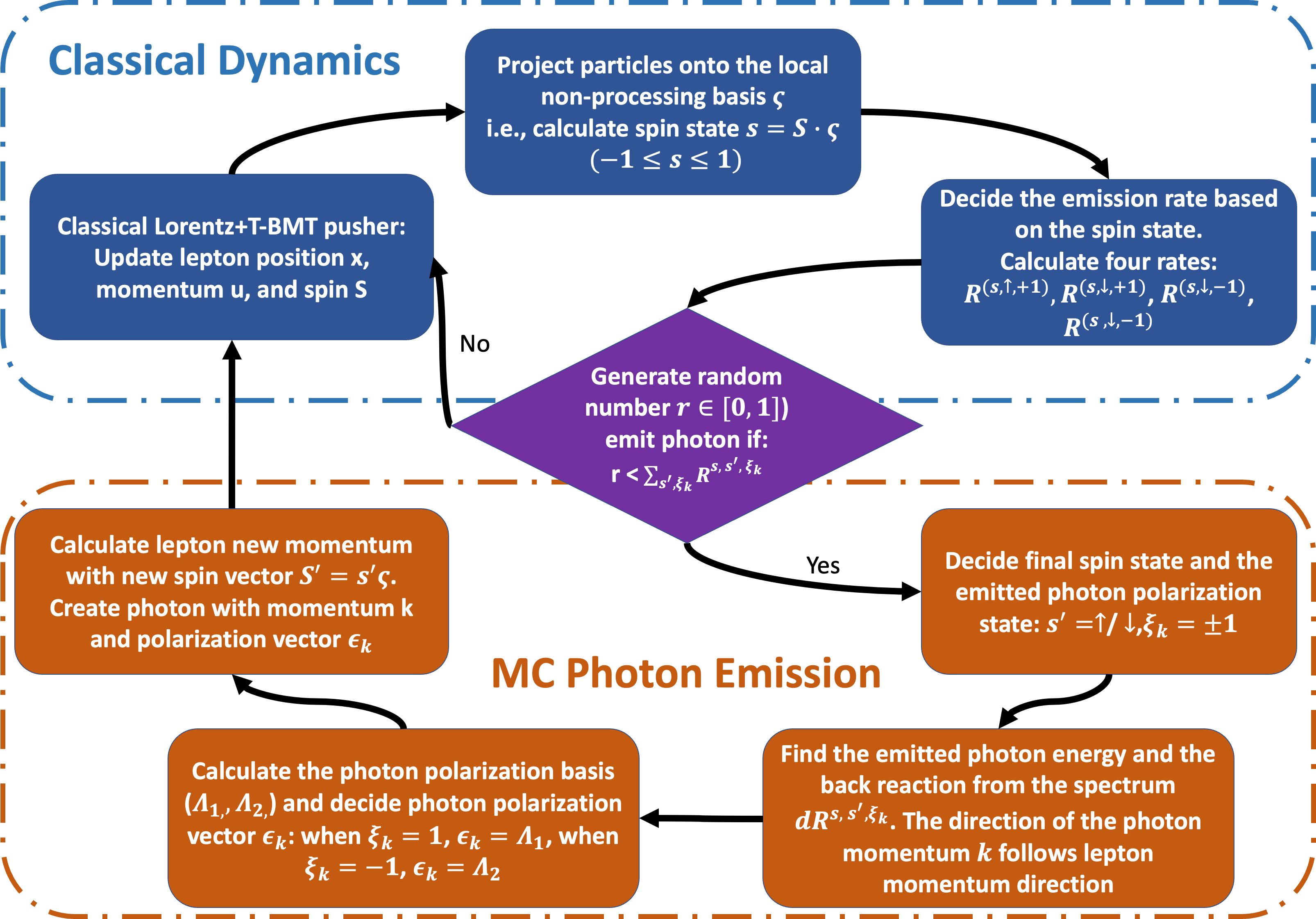}
\centering
\caption{Spin and polarization-involved quantum radiation module flow chart.}
\label{fig:QED_rad_flowchart}
\end{figure*}

The flow chart of spin and polarization-involved quantum radiation process calculation is shown in Fig.~\ref{fig:QED_rad_flowchart}. The code uses the classical Lorentz pusher and T-BMT equation-based spin pusher to update particle position, momentum, and spin for each time step. Then, we calculate the lepton's local non-precessing basis $\pmb\zeta$ and project the spin vector onto the basis to obtain the lepton spin state $s = \pmb S\cdot \pmb\zeta$. We also calculate the quantum parameter $\chi_q$ for each lepton. The particle state information $s$ and $\chi_q$ enable us to find the probability of the quantum radiation process. This probability is the criterion for entering the Monte Carlo-based spin and polarization-dependent quantum radiation module. In this module, we first determine the final spin state of the lepton $s'$ and the radiated photon Stokes parameter $\xi_k$. Then the photon's energy is obtained based on the radiation spectrum $R^{s, s', \xi_k}$ decided by the lepton's initial and final spin state and the Stokes parameter of the radiated photon. Due to the assumption of collinear emission, the direction of the photon momentum $\pmb k$ follows the lepton momentum direction. We calculate the polarization basis ($\pmb \Lambda_1$, $\pmb\Lambda_2$) and decide the direction of photon polarization vector $\pmb \epsilon_k$ based on $\xi_k$: if $\xi_k = 1$, $\pmb \epsilon_k = \pmb\Lambda_1$; if $\xi_k = -1$, $\pmb \epsilon_k =\pmb \Lambda_2$. Finally, we update the lepton momentum and the spin vector $\pmb S'$, which $\pmb S' = s' \pmb \zeta$.

\begin{figure*}
\includegraphics[width=15cm]
{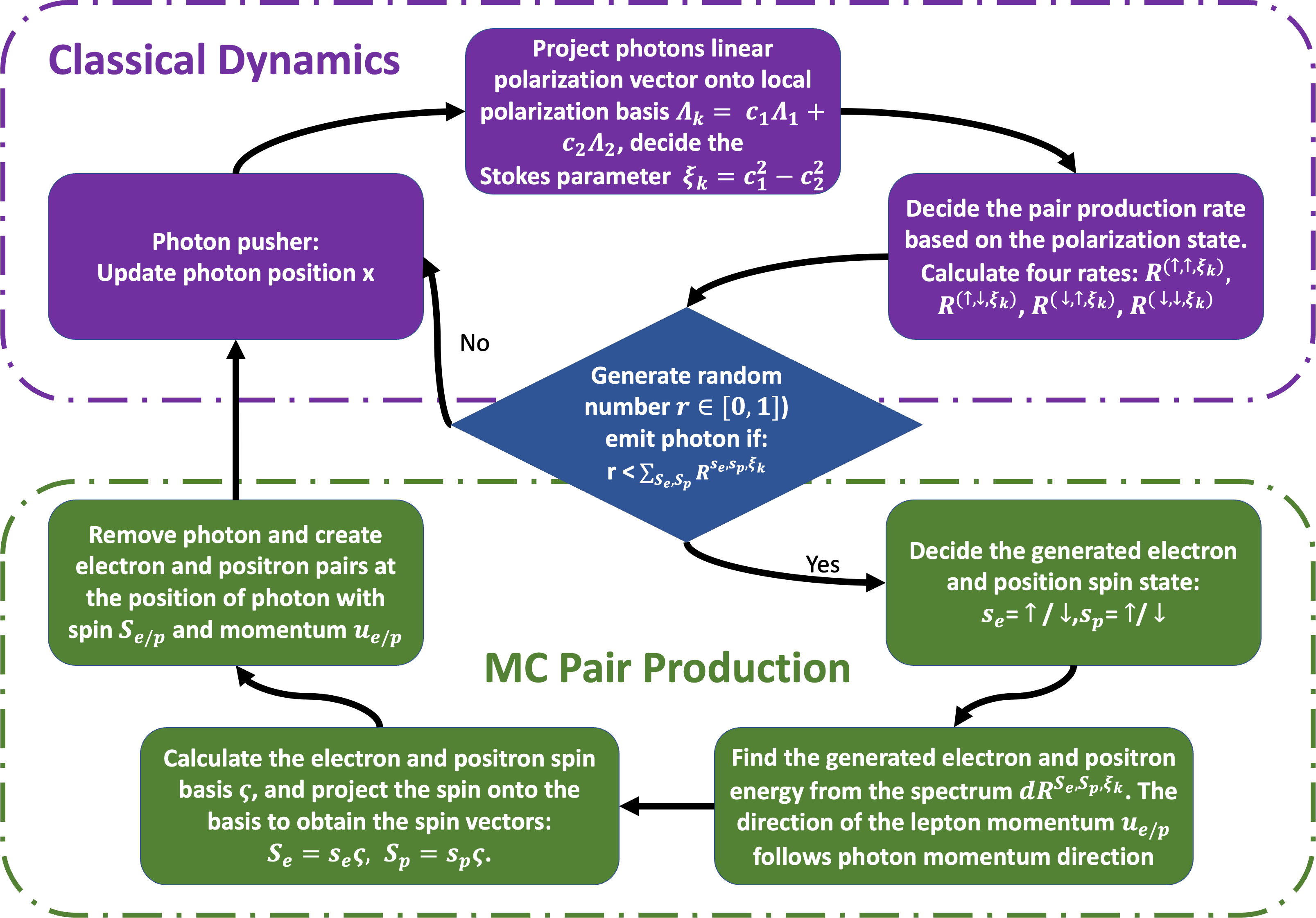}
\centering
\caption{Spin and polarization-involved pair production module flow chart.}
\label{fig:QED_pair_flowchart}
\end{figure*}
Fig.~\ref{fig:QED_pair_flowchart} is the flow chart for calculating the spin and polarization-involved pair production process. We begin with the classical photon pusher, which considers the photon as a particle traveling with the speed of light without any interaction in the medium. When a quantum process occurs, we calculate the polarization basis for the photon and project the photon polarization vector onto the basis to obtain the Stokes parameter $\xi_k$. The code uses the photon quantum parameter $\chi_q$ and its Stokes parameter $\xi_k$ to calculate the probability of polarization-dependent pair production. This probability is the criterion for entering the Monte Carlo-based spin and polarization-dependent pair production module. In this module, we first decide the generated electron and positron pair spin states $s_e$ and $s_p$. Then, we obtain the energy of the generated electron-positron pair based on the NBW spectrum $R^{s_e, s_p, \xi_k}$. The generated electron and positron pair momentum direction follows the photon momentum direction. We calculate the spin basis $\pmb \zeta$ for the generated pair to obtain their spin vector $\pmb S_{e/p} = s_{e/p} \pmb \zeta$. The photon that participates in the pair production process gives all its energy to the electron-positron pair and will be eliminated from the code.
{To benchmark the code performance, we reproduce the results Ref.~\onlinecite{Seipt_NJP_2021}, which is a sensitive test involving the interplay between the quantum emission rates and particle kinetics. This is shown in Appendix \ref{Appendix_test}.}

\section{Two pulse pair production}
In Sec.~\ref{Sec_background_spin_qed}, we show that the NBW process is asymmetric for  different photon polarization states. This effect can be illustrated by an all-optical experiment in which a high-energy electron beam from laser-wakefield acceleration collides with two linearly polarized laser pulses \cite{Wan_PRR_2020}. The schematic plot of this setup is shown in Fig.\ref{two_pulse_1} (a), which is similar to the setup used to probe vacuum
birefringence 
\cite{King_PRA_2016_birefringence, Nakamiya_PRD_2017_birefringence}. In the first collision, the linearly polarized laser pulse is designed to have a relatively long duration, to fully slow down the electron beam, and moderate intensity, to suppress the NBW process. The energetic, linearly polarized gamma rays from the first collision interact with the second laser pulse, which is more intense than the first pulse and can generate electron-positron pairs through the NBW process. The original electron beam, on the other hand, loses most of its energy in the first interaction, significantly reducing its capability to create electron-positron pairs when interacting with the second laser pulse. As a result, the electron-positron pairs generated from the two-pulse collision predominantly come from the interaction between the linearly polarized gamma-rays and the short, intense, second laser pulse. The relative polarization state of the gamma photon in the second pulse interaction can be simply controlled by the polarization direction relationship between the first and second laser pulses. The difference in yield will become maximum when two laser polarization directions change from parallel to perpendicular to each other. Notice that the setup proposed in Ref. \onlinecite{Wan_PRR_2020} uses a magnetic field to eliminate the leptons from the gamma-ray photons, which could remove the requirement for a long-duration first pulse. However, it will result in the interaction points between the first and second collisions being much farther away. More importantly, this increased distance could result in the gamma-ray generated from the first collision diverging significantly before interacting with the second pulse and, as a result, reducing the positron yield. Here, we use radiation force from beam-laser collision to stop electrons instead of deflecting the electron using magnetics. The separation between the two collision points could be minimized to reduce the influence of gamma-ray spreading, and the setup could be more compact.

\subsection{Scheme demonstration using polarization-dependent QED PIC simulations}

\begin{figure}[h]
\includegraphics{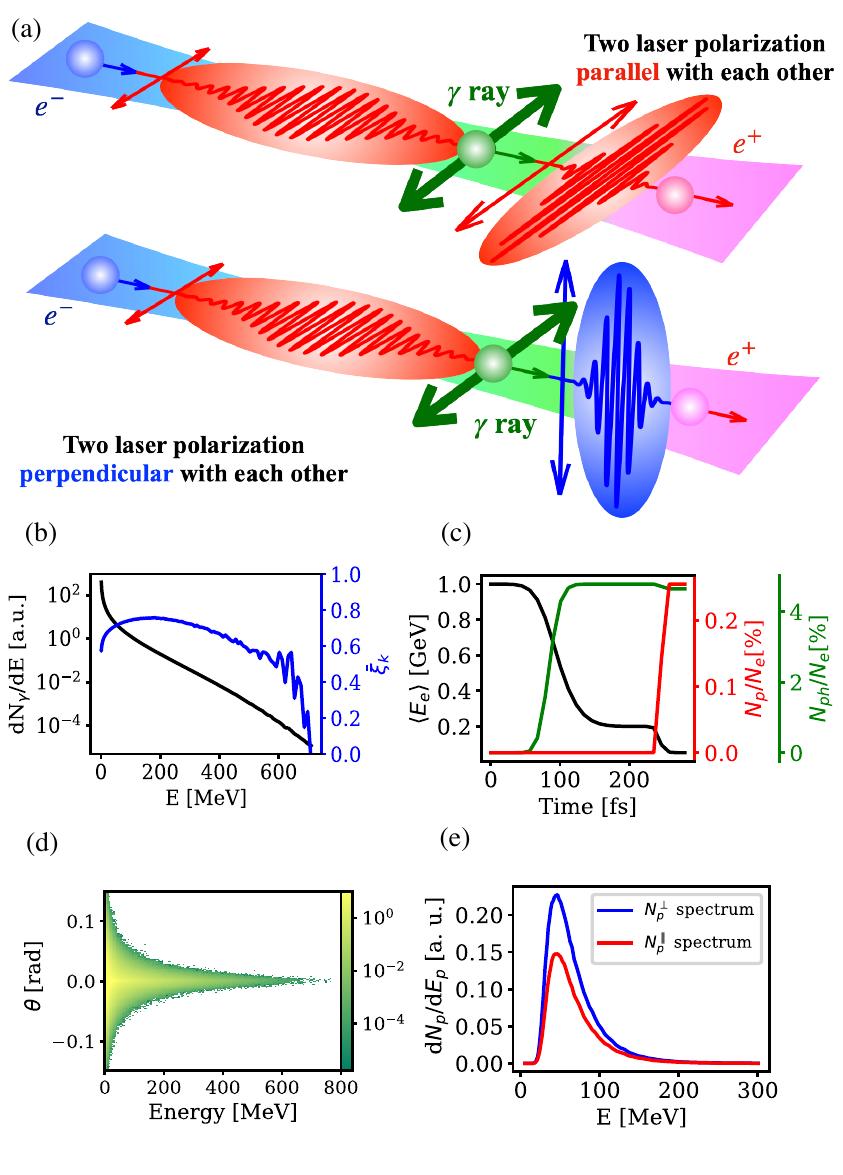}
\centering
\caption{(a) Two-pulse pair production setup. (b) Energy spectrum and averaged linear polarization stokes parameter of the gamma-ray beam from the first collision. (c) Temporal evolutions of the electron mean energy, positron yield, and high energy photon ($E>300$ MeV) yield.
  (d) Angular spectrum of the gamma-ray after the first collision. (e) The energy spectrum of the generated positron when the polarization direction of the two colliding beams is parallel or perpendicular. }
\label{two_pulse_1}
\end{figure}

 We conduct a pair of 2D simulations with our spin and polarization-dependent QED module to demonstrate this idea. In these two simulations, all the parameters remain the same, including the random seed in the QED Monte Carlo algorithm. The only thing we change is the polarization direction of the second laser, from parallel to perpendicular relative to the first laser. This restriction guarantees that the dynamics of the electron beam and the generated gamma-ray in the first collision are the same for both simulations, and the difference in the generated pairs can only come from the polarization effect in the NBW process. The first laser pulse in the collision has a peak $a_0$ = 30 and a duration of $t_{fwhm}$ = 120 fs. The second laser pulse follows right after the first pulse, with a peak $a_0$ = 160 and a duration of $t_{fwhm}$ = 19 fs. Both lasers have the same wavelength  $\lambda_0 = 0.8\ \mu m$, and focal spot radius $w_0 = 10\lambda_0$. The electron beam in the simulations starts unpolarized with an average energy of $\langle E\rangle = 1$ GeV, relative energy spread $\sigma_E/\langle E\rangle=10\%$, beam length $l_e$ = $2\lambda_0$, beam radius $w_e$ = $2\lambda_0$, density $n_e \approx 10^{18}$ cm$^{-3}$ with a transversely and longitudinally Gaussian distribution, which is achievable for current laser wakefield acceleration (LWFA) technology. Each macro-particle in the simulations starts with zero spin vector length to represent the initially unpolarized spin state. In the quantum calculation, the probability of these particles being in a spin-up or spin-down state relative to the basis is equal. Fig.~\ref{two_pulse_1} (b)-(e) shows the results of the simulations. The first collision generates an energetic gamma-ray with a maximum linear polarization degree that reaches over $\xi_k = 70\%$, as shown in Fig.~\ref{two_pulse_1} (b). Fig.~\ref{two_pulse_1} (c) shows the temporal evolution of the whole interaction process. The mean energy of the electron beam reduces to below 200 MeV in the first collision; this corresponds to a maximum quantum parameter $\chi_e<0.27$ when interacting with the second laser. The energy of the electron beam is converted into high-energy photons in the first collision. They will then collide with the second laser and generate positrons in the second collision. The green curve in Fig.~\ref{two_pulse_1} (c) plots the number of photons above 300 MeV, which increase in the first collision and decrease in the second collision. The red curve plots the positron number, which only increases during the second collision.
 The generated gamma-ray beam has a divergence of about 5 $mrad$ for energy larger than 300 MeV (Fig.~\ref{two_pulse_1} (d)).  Fig.~\ref{two_pulse_1} (e) shows the energy spectrum of the positron generated from the second collision. When we change the polarization direction of the second laser pulse from parallel to perpendicular to the first pulse, we find the difference of the positron yields $\Delta_p = \frac{N_p^{\perp}-N_p^{\parallel}}{N_p^{\parallel}}\approx 50\%$. The total charge of the generated positron beam is about $0.002$ of the initial electron beam. For the GeV class electron beam generated by LWFA, the beam charge is $\sim 10$~pC. There will be over $10^5$ positron generated from the collision, which can give good statistics to illustrate this effect in an actual experiment. 

\subsection{Analytic model for polarization dependent pair production yield}
To help understand the scaling of the electron-positron pair creation, we first formulate scaled equations describing the photon emission, electron beam energy loss, and pair creation in the second laser. For simplicity, we assume the two lasers to be square pulses with infinite spot sizes. This reduces the problem to one dimension. The model does not include 3D effects, such as the overlapping between the laser spot and the electron beam, which could be important for estimating the positron yield. The polarization effect induced positron yield difference $\Delta_p$ is less sensitive to the 3D effects and should be well captured by our simplified 1D model. We start from the equation for the radiative energy loss \cite{Ridgers_JPP_2017}:

\begin{equation}
    \frac{d\gamma}{dt} = -\frac{2\alpha_f c}{3\lambdabar_c}\chi_e^2(t)g(\chi_e)\;,
\label{radiation_reaction_equation}
\end{equation}

where $g$ is the Gaunt factor for quantum radiation correction, $g(\chi_e)\approx[1 + 4.8(1 + \chi_e)\rm{ln}(1 + 1.7\chi_e) + 2.44\chi_e^2]^{-2/3}$. We define scaled time $\tau$ and scaled energy $\Gamma$ with respect to the initial energy $\gamma_0$:
\begin{equation}
 \tau =\gamma_0 t, \ \Gamma(\tau) = \frac{\gamma}{\gamma_0}\;.
\end{equation}

The equation for the radiation force can therefore be rewritten as:        


\begin{equation}
\frac{d\Gamma}{d\tau} = - \frac{2\alpha_f c}{3\lambdabar_c} \chi_1^2\Gamma^2g(\chi_1\Gamma)
\label{radiation_energy_reduction}
\end{equation}

Here, $\chi_{1} = 2\frac{\hbar\omega_0}{m_ec^2}a_{01}\gamma_{0}$ is the quantum parameter calculated using the \emph{initial} energy of the electron beam $\gamma_0$ and normalized intensity $a_{01}$ of the first laser,  $\chi_e = \chi_1\Gamma$. 




The solution to this equation is
\begin{equation}
\Gamma(\tau) =  \frac{1}{1+\frac{2\alpha_f c}{3\lambdabar_c} \chi_1^2\int_0^\tau d\tau' g(\chi_1\Gamma(\tau'))}
\label{Gamma_equation}
\end{equation}

Note that when $\chi_1\Gamma\ll1$, the Gaunt factor $g$  tends to 1. In this case, equation \ref{Gamma_equation} will recover the exact solution for classical radiation reaction: $\Gamma(\tau) = 1/\left(1+\frac{2\alpha_fc}{3\lambda_c}\chi_1^2\tau\right)$.

Now assume an electron beam starts with quantum parameter $\chi_1$ and interacts with a square laser pulse whose duration in the rest frame of the electron beam at the beginning is $\tau_1$ ($\gamma_0\tau_1$ in the lab frame). It will generate polarized gamma-rays with spectrum $S(\omega_\gamma, \xi_k)$, where $\omega_\gamma$ is the energy and $\xi_k$ is the polarization state of the gamma-ray beam; in our model $\xi_k \in \{-1, 1\}$.  This radiation spectrum can be obtained by integrating the polarization-resolved NLC spectrum generated at every time step over the total interaction time: 
\begin{equation}
S(\omega_\gamma, \xi_k) = \int_0^{\gamma_0\tau_1} dt \frac{F_{\text{NLC}}(\chi_e(t),\omega_\gamma,\xi_k)}{\gamma(t)}
\end{equation}

The explicit form of the polarization-resolved NLC spectrum $F_{\text{NLC}}(\chi_e, \omega_\gamma, \xi_k)$ is given in \ref{NLC_ave_diff_rate} in the appendix \ref{Appendix:Spin QED equations}, where we assume the electron beam to stay unpolarized during the interaction. Replace $\gamma$ and $t$ with scaled energy and time $\Gamma$, $\tau$, and we get: 
\begin{equation}
S(\chi_1,\tau_1,\omega_\gamma,\xi_k) =\int_0^{\tau_1} d\tau \frac{F_{\text{NLC}}^{\pi/\sigma}\left(\chi_1\Gamma(\tau),\omega_\gamma, \xi_k\right)}{\Gamma(\tau)}
\end{equation}

We then calculate the number of electron-positron pairs after the radiated photon interacts with the second square laser pulse. The second laser pulse's duration in the counter-propagating electron beam's rest frame with energy $\gamma_0$ is $\tau_2$. We assume that the electron beam has lost most of its energy after the first collision, so it won't create pairs in the second collision. Hence, the number of pairs generated from the second collision only depends on polarization-resolved NBW rate $R_{\text{NBW}}$ and the incoming gamma ray spectrum $S$, we integrate them over the total interaction time and also the energy of the gamma-ray participates in the pair production process:

\begin{equation}
\begin{split}
&N_{\text{pair}} (\chi_1,\tau_1,\chi_2,\tau_2)= \\
&\sum_{\xi_k=\pm1}\int_0^\infty d\chi_{\gamma}\int_0^{\tau_2}d\tau\frac{R_{\text{NBW}}(\chi_{\gamma}, \xi_k^o)S(\chi_1,\tau_1,\omega_\gamma,\xi_k)}{\chi_{\gamma}/\chi_2}
\end{split}
\label{normalized_pair_generation}
\end{equation}

Here, the explicit form of the polarization-resolved NBW rate $R_{\text{NBW}}(\chi_{\gamma}, \xi_k^o)$ is given in \ref{NBW_ave_rate} in the appendix \ref{Appendix:Spin QED equations}. Note that $\xi_k^o$ is the stokes parameter in the observation frame of the second laser, which for the polarization direction of the second laser to be either parallel or perpendicular to the first laser, $\xi_k^o = \pm \xi_k$. $\chi_2 = 2\frac{\hbar\omega_0}{m_ec^2}a_{02}\gamma_{0}$ is the quantum parameter calculated using the initial energy of the electron beam $\gamma_0$ and normalized intensity $a_{02}$ of the second laser. $\chi_{\gamma} = \chi_2\frac{\omega_\gamma}{\gamma_0}$ is the quantum parameter of the photon in the second laser field. Note that this  set of equations shows that the scaled radiation reaction and pair creation dynamics does not explicitly depend on the laser parameters $a_{0,i}$ and $T_i$ and initial beam energy $\gamma_0$ independently, but only on $\chi_i\propto a_{0,i}\gamma_0$ and $\tau_i=T_i/\gamma_0$, where $i=1,2$, in addition to their relative polarization.

\subsection{Designing  optimal parameters.}

\begin{figure}[h]
\includegraphics{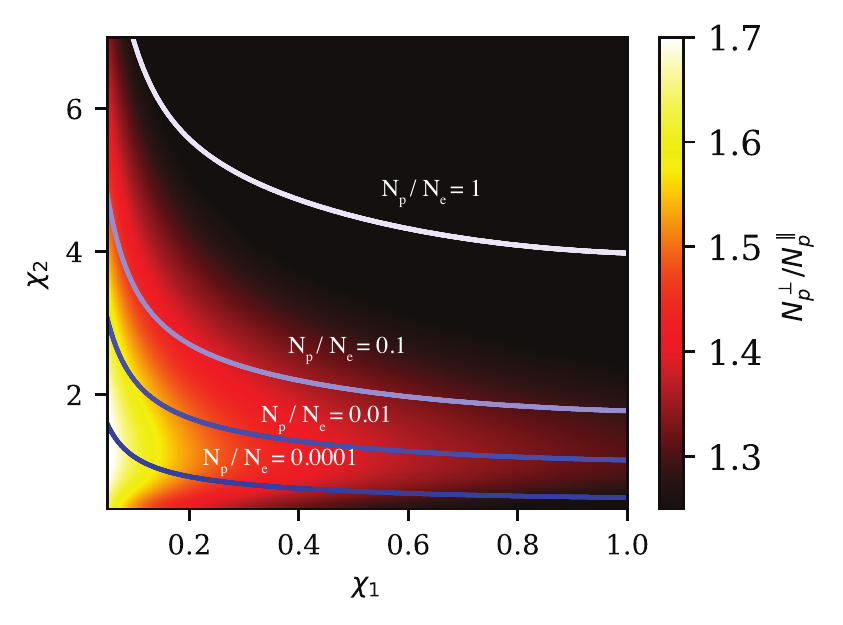}
\centering
\caption{The phase space plot for the ratio of the positron yield when changing the polarization directions of two colliding laser pulses from parallel to perpendicular for different $\chi_1$ and $\chi_2$.}
\label{TPP_phase_1}
\end{figure}

Now we show explicitly that the polarization-resolved electron-positron pair generation $N_{\text{pair}}$ effectively only depends on parameters $\chi_1$, $\chi_2$, $\tau_1$, $\tau_2$ and also on the relative polarization of two lasers. Based on equation \ref{normalized_pair_generation}, we generate the phase space plot Fig. \ref{TPP_phase_1} for the ratio of the positron yield when changing the relative polarization directions of the two colliding laser pulses from parallel to perpendicular for different $\chi_1$ 
(from 0.05 to 1.0) and $\chi_2$ (from 0.5 to 7.0). The first laser pulse duration is set to be long enough to reduce the electron energy so that it won't create many pairs when interacting with the second laser. The second laser pulse scaled time duration $\tau_2 = T_2/\gamma_0 = 0.027$ fs. The contour on the plot shows the number of positrons generated compared with the initial electrons. We verified our numerical calculation results with simulations using the spin and polarization-dependent QED code. The simulations' parameters and their results are collected in table \ref{TPP_sim_table} and compared with the predictions given by the model. We find our model well predicts the polarization effect in the positron yield. The absolute positron yield predicted by the model is always smaller than the PIC simulation result but of the same order of magnitude. The model provides a reasonable estimation of the actual PIC simulation. The discrepancy in positron yield probably comes from the fact that the model does not include the impact of stochastic in the calculation, which can underestimate the number of high-energy photons generated from quantum radiation.
From the phase space plot and the table, we can find a maximum difference of over 70\% is achievable using this setup. However, there is also a trade-off between generating more positrons and increasing the difference in yield when we change the laser polarization direction. We can find that the region in which most positron generated is also where the difference in yield is below 30\%. The region in which the difference is highest is where the positron charge is less than $10^{-4}$  of the initial electron beam charge. We can also find that lower $\chi_1$ can result in a higher difference in positron yield when the laser's polarization direction changes. The reason for this is that the NLC process at lower $\chi_1$ can generate gamma-rays with a higher linear polarization $\xi_k$. The NBW process's polarization response can explain why the positron yield difference is higher when $\chi_2$ is between 1 and 2. 


\begin{table*}[t]
  \centering
\begin{tabular}{c c c c c c c c c}
 $\ \ \chi_1\ \ $  & $\ \tau_1$ [fs] &  $\ \ \chi_2\ \ $ &  $\ \tau_2$ [fs] & $\ $ Mean energy [GeV] $\ $ & $\  \left(N_p^{\parallel}/N_e\right)^{sim} \ $ & $\ 
\left(N_p^{\parallel}/N_e\right)^{model} \ $ & $\ $ $\Delta_p^{sim}$  $\ $ & $\ $ $\Delta_p^{model}$  $\ $ \\
\hline
0.47 & 0.164  & 3.55  & 0.027 & 1 & 0.56 & 0.45 & 23\% & 26\%\\ 

0.236 & 0.164  &  2.84  & 0.027 & 1 & 0.15 & 0.11 & 38\% & 40\% \\

0.236 & 0.164  & 1.42  & 0.027 & 1  &0.0053 & 0.0034  & 49\% & 49\%\\

0.236 & 0.164  & 1.42  & 0.027 & 2 &0.0053 & 0.0034 & 49\% & 49\%\\

0.236 & 0.164 & 1.42  & 0.027 & 4  &0.0054 & 0.0034 & 50\% & 49\%\\

0.118 & 0.327 & 1.42  & 0.027 & 1 & $8.5\times 10^{-4}$ & $6\times 10^{-4}$ & 60\%  & 59\% \\

0.05 & 0.654  & 1.42  & 0.027 &  1 &  $2.85\times 10^{-5}$ &  $2.3\times 10^{-5}$ & 71\% & 72\%

\end{tabular}

\caption{Simulation parameters used to investigate the optimized parameters to maximize the measureable difference in positron yield when we change the laser polarization direction and their results. Here the pulse duration is defined in scaled time $\tau = t/\gamma_0$. We compare the positron yields, measurement by the ratio between the number of generated positrons and initial beam charge, and the polarization effect induced difference in positron yields $\Delta_p = (N_p^{\perp}-N_p^{\parallel})/N_p^{\parallel}$ in the simulations with the predictions from our model.}
\label{TPP_sim_table}
\end{table*}

\begin{figure}[h]
\includegraphics{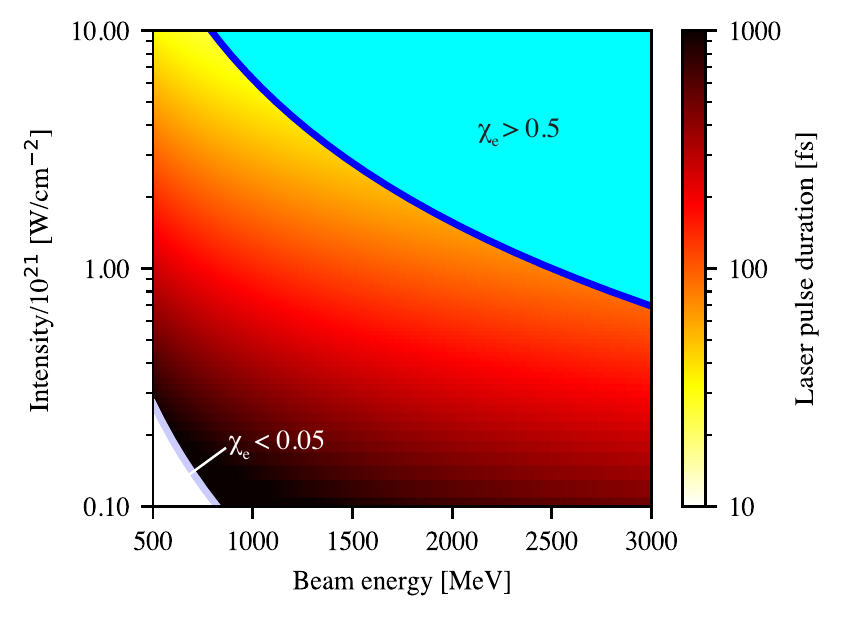}
\centering
\caption{The minimum requirement for pulse duration of the first laser pulse for different input beam energies and laser intensities in the first collision. The regions on the upper right and lower left are where $\chi_e<0.05$ and $\chi_e>0.5$.}
\label{TPP_laser_requirement_1}
\end{figure}

We have shown how the difference in positron yield depends on normalized parameters $\chi_1$, $\chi_2$, $\tau_1$, $\tau_2$. To design an actual experiment, giving some criteria for the laser and electron beam parameters under practical units are necessary. We start by considering the criteria for the first laser pulse. We want the pulse duration to be long enough to reduce the energy of the electron beam to the level that its interaction with the second laser pulse in the NBW pair production process becomes unimportant. Using equation \ref{radiation_reaction_equation}, we predict the minimum requirement of the pulse duration to reduce the electron beam mean energy below 10\% of its original value for different laser peak intensities and electron beam initial energy in Fig.~\ref{TPP_laser_requirement_1}. The reason to choose 10\% initial beam energy as the energy reduction criteria is because, as shown in Eq.~\ref{radiation_energy_reduction}, the ability of radiation reaction to slow down the electron beam at this energy level is much weaker than at the beginning. The white and cyan color regions in Fig.~\ref{TPP_laser_requirement_1} marks the space unsuitable for the experiment; we want the maximum quantum parameter $\chi_{e}(t) = 2\frac{\hbar\omega_0}{m_ec^2}a_{0}
\langle\gamma_{e}\rangle_{t=0}$  to be within the range of 0.05 and 0.5. The white region at the bottom left corner of the plot marks the parameter space for $\chi_e<0.05$, in which insufficient high-energy photons will be generated. The cyan color region at the upper right corner marks the parameter space $\chi_e>0.5$, in which the NBW pair production process is substantial when the electron beam interacts with the first laser pulse. The generated pairs from the first collision, {contributing to the background,} will reduce the difference in the positron yield when changing the laser polarization direction in the second collision.

\begin{figure}
\includegraphics{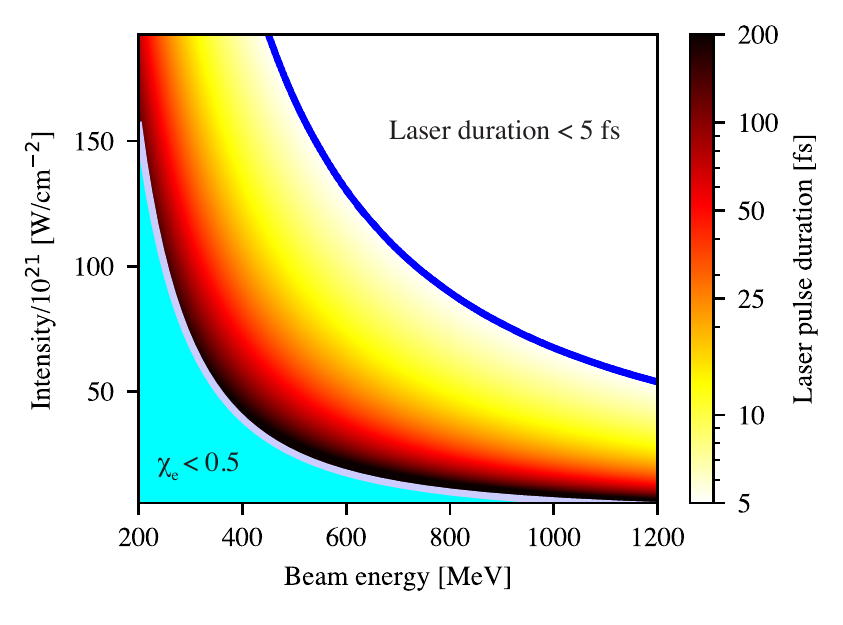}
\centering
\caption{The maximum duration of the second laser pulse  for different input gamma-ray beam energies and laser intensities in the second collision. The regions on the lower left are where $\chi_e<0.5$. The upper right corner is the region with laser duration is $t_2< 5$ fs.}
\label{TPP_laser_requirement_2}
\end{figure}


In designing the parameters for the second laser, we also need to consider the depletion of the high-energy photons by the NBW process when the pulse duration is too long. Suppose that all high-energy photons turn into electron-positron pairs after the second collision. In this case, we cannot observe a difference in positron yield when we change the laser polarization direction. Using equation \ref{deplection effect prediction} and presume the input photon beam to have an averaged stokes parameter of the input photon beam  $\bar\xi_k = 0.6$ ($N_\gamma^{\xi_k=1} = 4N_\gamma^{\xi_k=-1}$), we generate Fig.~\ref{TPP_laser_requirement_2} that predicts the maximum laser duration we can have for which greater than 25\% difference in positron yield is observable when changing the laser pulse polarization direction for varying laser peak intensity and input beam mean energy: 
\begin{equation}
\begin{split}
&\Delta_p = \frac{N_p^{\perp}-N_p^{\parallel}}{N_p^{\parallel}}\\ &=\frac{\left(N^{\xi_k=1}_\gamma- N^{\xi_k=-1}_\gamma\right)\left(e^{-R_{\text{NBW}}^{\xi_k=1}t}-e^{-R_{\text{NBW}}^{\xi_k=-1}t}\right)}{N^{\xi_k=1}_\gamma\left(1-e^{-R_{\text{NBW}}^{\xi_k=1}t}\right)+N^{\xi_k=-1}_\gamma\left(1-e^{-R_{\text{NBW}}^{\xi_k=-1}t}\right)}
\end{split}
\label{deplection effect prediction}
\end{equation}
In Fig.~\ref{TPP_laser_requirement_2}, we mark two regions that are not ideal for the experiment. The upper right corner of the plot shows a pulse duration smaller than $5 $~fs, which is too short to generate with current technology. The bottom left corner marks the region in which the maximum quantum parameter $\chi_{\gamma}= 2\frac{\hbar\omega_0}{m_ec^2}a_{0}\langle\gamma\rangle_{t=0}$ is smaller than $0.5$. Here, we consider the resolution of observing the difference in positron yield $\Delta_p$ when we change the laser polarization direction, which can be estimated through the statistical uncertainty  $1/\Delta_p\sqrt{N_p}$. For a single shot measurement, assuming $\Delta_p=0.5$; having more than $\sim 1000$ positrons generated will be desirable. This corresponds to a resolution of about 5\%. A laser wakefield beam typically contains charge on the order of several pC, which contains about $10^7$ electrons. The positron yield should be at least $10^{-4}$ of the initial electron beam charge. According to Fig.~\ref{TPP_phase_1}, when $\chi_2 < 0.5$, the positron yield will always be below $10^{-4}$ of the initial electron beam charge. Below this threshold, there won’t be many positrons generated during the interaction, which makes the positron charge measurement unlikely.

The parametric study of the polarization dependence NBW process under two pulses pair production setup provides insight into the ideal parameters to observe a clear signal of polarization effect in pair production yield for an all-optical experiment. Here, we propose a set of optimal experimental parameters. For a 1 GeV electron beam from the laser wakefield accelerator, the laser pulse for the first collision needs to have a moderate intensity of $0.2 - 1.5 \times 10^{21}$ [W/cm$^{-2}$] and a long duration of 0.1 to 1 ps. The second laser pulse should be with an intensity of $10 - 90 \times 10^{21}$ [W/cm$^{-2}$] and a duration of $20-40$ fs. In the ideal scenario, a $40 - 60\%$ difference in positron yield is expected with the proposed parameter when we change the laser polarization direction. 

\section{Conclusion}

With experimental studies of the strong-field QED regime in the laboratory likely to be realized with new facilities coming online, developing a more accurate QED module with the effect of lepton spin and photon polarization taken into account becomes necessary. Recent studies of how including spin and polarization in the calculation will result in a considerable difference in QED cascade simulation \cite{Seipt_NJP_2021, Seipt_PRA_2020, Song_NJP_2021}, as well as polarized gamma-ray and lepton beams generation through strong field QED process\cite{Seipt_PRA_2019, Li_PRL_2019, Li_PRL_2020_gamma_ray, Li_PRL_2020_helicity_transfer}, add to the value of developing a full spin and polarization-resolved QED module in PIC code. This work presents our spin and polarization-resolved quantum radiation reaction module based on PIC code framework Osiris 4.0. The success of reproducing Ref.~\onlinecite{Seipt_NJP_2021}'s main results of studying the polarized seeded cascade marks the code's reliability in dealing with complicated multi-stage QED processes. We have used this to demonstrate a two-pulse-pair production scheme for experimentally measuring the effect of the gamma-ray polarization state on the NBW pair creation and find the optimized condition for maximizing the yield of pair production when we rotate the laser polarization direction. This was achieved through our numerical model and parameterized through a set of normalized differential equations. The  simulation result predicts a difference in yield of over 50\% by simply changing the polarization directions of two linearly polarized laser pulses, which should be an easily measurable signature in a real experiment. We also broadly discuss the criteria for the
laser and electron beam parameters under practical units to design an actual experiment, which is achievable in the near future. Notice that the model we present has the limitation that it requires the lepton beam to be initially unpolarized and using a field with a fixed direction, like a linearly polarized laser\cite{Chen_PRD_2022}. As a result, our simplified model applies to the situations studied in this work. In a general field configuration, the lepton spin and photon polarization components combining all three orthogonal directions ($\boldsymbol\varepsilon$, ${\pmb{\beta}}$, $\pmb{\kappa}$) must be considered and is currently being implemented in the Osiris 4.0 framework.

\appendix 
\section{Cross section of spin and polarization-resolved QED process}\label{Appendix:Spin QED equations}
The spectrum of spin and polarization-resolved NLC process we use in the model follows equation 38 in Ref. \onlinecite{Seipt_PRA_2020}:

\begin{equation}
\begin{split}
{F}_{\text{NLC}}(s, & s',\xi_k)  =  -\frac{\alpha}{4b_p}\biggl\{[1+ss'+\xi_kss'(1-g)]\mathrm{Ai}_1(z) \\
& +\left[\lambda s + \frac{\lambda}{1-\lambda}s' + \xi_k\left(\frac{\lambda}{1-\lambda}s+\lambda s'\right)\right]\frac{\mathrm{Ai}(z)}{\sqrt{z}} \\
& +\left(g+ss'+\xi_k\frac{1+gss'}{2}\right)\frac{\mathrm{Ai}'(z)}{z}\biggl\}
\end{split}
\label{NLC_diff_rate}
\end{equation}

Here, $\alpha = e^2/4\pi$ is the fine structure constant. $b_p = p\cdot \kappa/m^2$ is the quantum energy parameter, which $p$ is the momentum of the lepton before radiation, and $\kappa$ is the wave vector of the colliding laser. $\mathrm{Ai}$ is the Airy function, and $\mathrm{Ai}'$, $\mathrm{Ai}_1$ are it's derivative and integral. The argument of the Airy function $z = (\frac{\lambda}{\chi_e(1-\lambda)})^{2/3}$ depends on the quantum parameter of the lepton $\chi_e$. $\lambda = \frac{p^-_{out}}{p^-_{in}}$ is the normalized light-front momentum transfer, which under the condition of head-on collision configuration with the incoming particle highly relativistic, can be approximated as $ \lambda \simeq \omega_\gamma/\varepsilon$, where $\omega_\gamma$ is the emitted photon energy and $\varepsilon$ the lepton energy before radiation. $g = 1+ \lambda^2/[2(1-\lambda)]$. 

For the case of the photon polarization-resolved NLC process for unpolarized electrons, the spectrum can be achieved by setting $s = s' = 0$ in Eqn. \ref{NLC_diff_rate} and multiplying the whole equation by 2. This is equivalent to performing an average over the initial spin and sum over the final spin states of the lepton:

\begin{equation}
\begin{split}
{F}_{\text{NLC}}^{\text{unpol}} & (\xi_k)   \\ & =  -\frac{\alpha}{2b_p}\left[\mathrm{Ai}_1(z)+(2g+\xi_k)\frac{\mathrm{Ai}'(z)}{z}\right]
\end{split}
\label{NLC_ave_diff_rate}
\end{equation}

The polarized NBW pair spectrum comes from equation 59 in Ref. \onlinecite{Seipt_PRA_2020}.

\begin{equation}
\begin{split}
F_{\text{NBW}} & (s_e,  s_p,\xi_k)  \\ & =  -\frac{\alpha}{4b_k}\biggl\{[1+s_es_p+\xi_ks_es_p(1-\tilde g)]\mathrm{Ai}_1(\tilde z) \\
& +\left[\frac{s_e}{\lambda} - \frac{s_p}{1-\lambda} + \xi_k\left(\frac{s_p}{\lambda}-\frac{s_e}{1-\lambda}\right)\right]\frac{\mathrm{Ai}(\tilde z)}{\sqrt{\tilde z}} \\
&+\left(\tilde g+s_es_p+\xi_k\frac{1+\tilde gs_es_p}{2}\right)\frac{2\mathrm{Ai}'(\tilde z)}{\tilde z}\biggl\}
\end{split}
\label{NBW_diff_rate}
\end{equation}


Here, the argument of the Airy function $\tilde z = [\chi_\gamma\lambda(1-\lambda)]^{-2/3}$ depends on the photon quantum parameter $\chi_\gamma$. The quantum energy parameter $b_k = k\cdot \kappa/m^2$ is related to the center-of-mass energy of the incident photon colliding with the plane-wave laser field, with k being the momentum of the photon. $\lambda \simeq \varepsilon_p/\omega_\gamma$ which $\varepsilon_p$ is the energy of the generated positron, $\omega_\gamma$ is the energy of the incident photon, $\tilde g = 1- \frac{1}{2\lambda(1-\lambda)}$.

For the case of a polarized photon decay into an unpolarized electron-positron pair, the spectrum can be achieved by setting $s_e = s_p = 0$ in Eqn. \ref{NBW_diff_rate} and multiplying the equation by 4:

\begin{equation}
\begin{split}
F_{\text{NBW}}^{\text{unpol}}   & (\xi_k)   \\ & =  -\frac{\alpha}{b_k}\left[\mathrm{Ai}_1(\tilde z)+(2\tilde g+\xi_k)\frac{\mathrm{Ai}'(\tilde z)}{\tilde z}\right]
\end{split}
\label{NBW_ave_diff_rate}
\end{equation}

Finally, the rate of a polarized photon decay into an unpolarized pair in the NBW process can be obtained by integrating NBW spectrum Eqn. \ref{NBW_ave_diff_rate} over $\lambda$:

\begin{equation}
\begin{split}
{R}_{\text{NBW}}^{\text{unpol}}  & (\xi_k)\\ & =  -\int_0^{1} \frac{\alpha}{b_k}\left[\mathrm{Ai}_1(\tilde z)+(2\tilde g+\xi_k)\frac{\mathrm{Ai}'(\tilde z)}{\tilde z}\right]d\lambda
\end{split}
\label{NBW_ave_rate}
\end{equation}

We can find that photon with polarization state $\xi_k = 1$ has a higher NBW pair production rate than  $\xi_k = -1$. 

\section{Benchmarking the particle-in-cell implementation}
\label{Appendix_test} To benchmark the code performance, we try to reproduce the result in the paper ``Polarized QED cascades'' \cite{Seipt_NJP_2021} using our spin and polarization-involved QED PIC. This paper studies the avalanche-type cascades, which could occur at the rotating electric fields at the magnetic nodes for two counter-propagating circularly polarized laser pulses. Such an avalanche-type cascade exhibits an exponential growth in particle number, limited by the available (laser) field energy. The paper discusses two different scenarios: lepton and seeded gamma-ray cascade. The polarized QED cascade process is complicated because the spin and polarization involved in NLC and NBW processes are strongly coupled. The NLC process polarized the lepton while generating linearly polarized gamma-ray. The gamma-ray's polarization state will influence the NBW pair production rate. At the same time, the generated electron-positron pair is also polarized, modifying lepton momentum and spin distribution. As a result, reproducing the polarized QED cascade result can comprehensively test the performance of our spin and polarization-involved QED module. Notice that the paper uses notation $\parallel$ and $\perp$ instead of $\pmb \sigma$ and $\pmb \pi$ for the photon polarization state, which has a similar meaning. 

\begin{figure}[h]
\includegraphics[width=8.0cm]{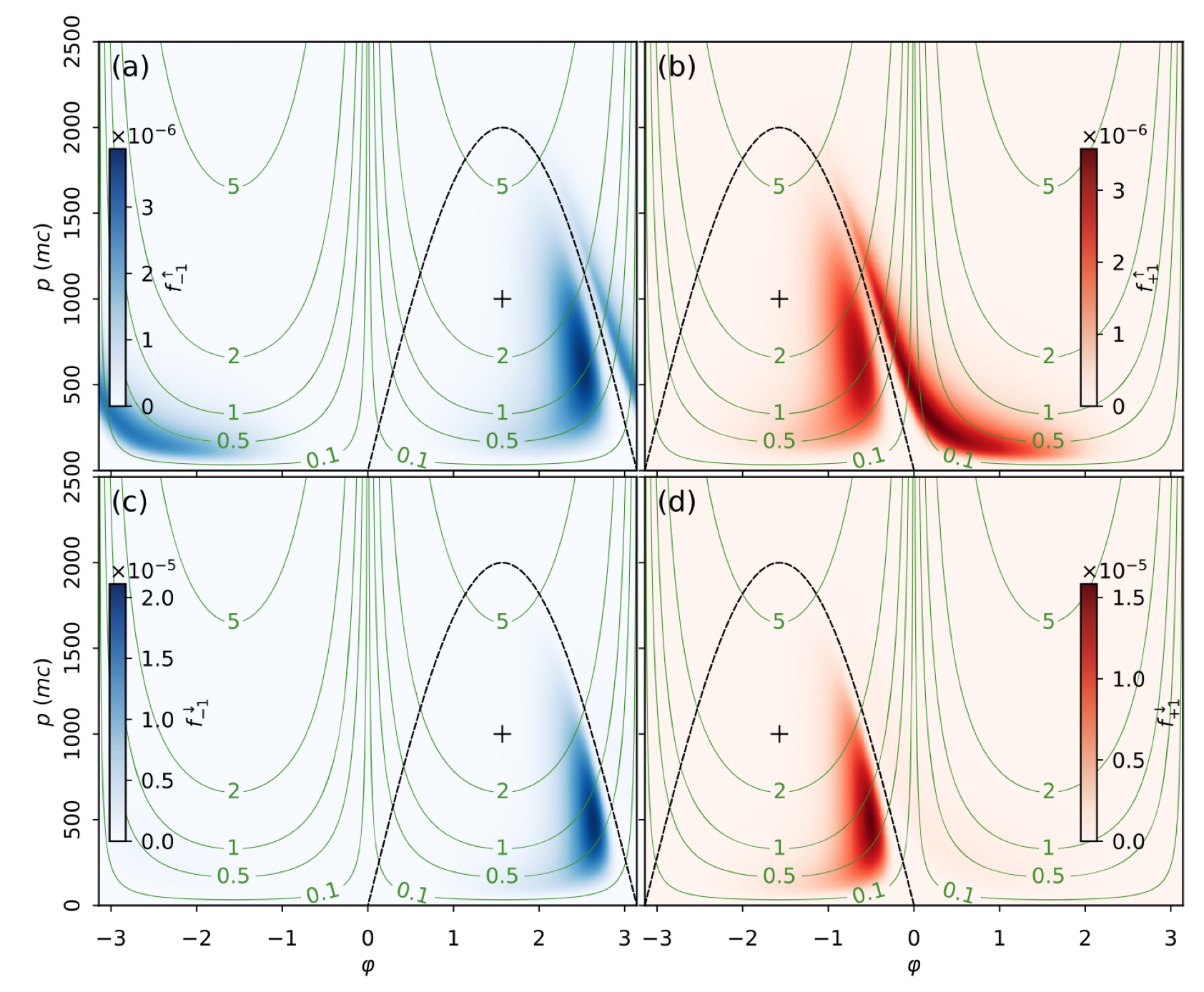}
\centering
\caption{Snapshot of the electron (a), (c) and positron (b), (d) distribution functions in an up (a), (b) or down (c), (d) spin state for $a_0 = 10^3$ and $\omega t = 10$ in a rotating radial frame. Green curves are $\chi$ isocontours. Black dashed curves represent the separatrix of the classical advection $p = -2a_0q \sin\varphi$, and black crosses are the corresponding fixed points at $\varphi = -q \pi/2.0$, $p = a_0$. This figure is from ``Polarized QED cascades'' New J. Phys. \textbf{23} 053025 (2022)  by D. Seipt, C. P. Ridgers, D. Del Sorbo, and A. G. R. Thomas, which is licensed under CC BY 4.0.}
\label{fig:phase_space}
\end{figure}

\begin{figure}[h]
\includegraphics[width=7.3cm]{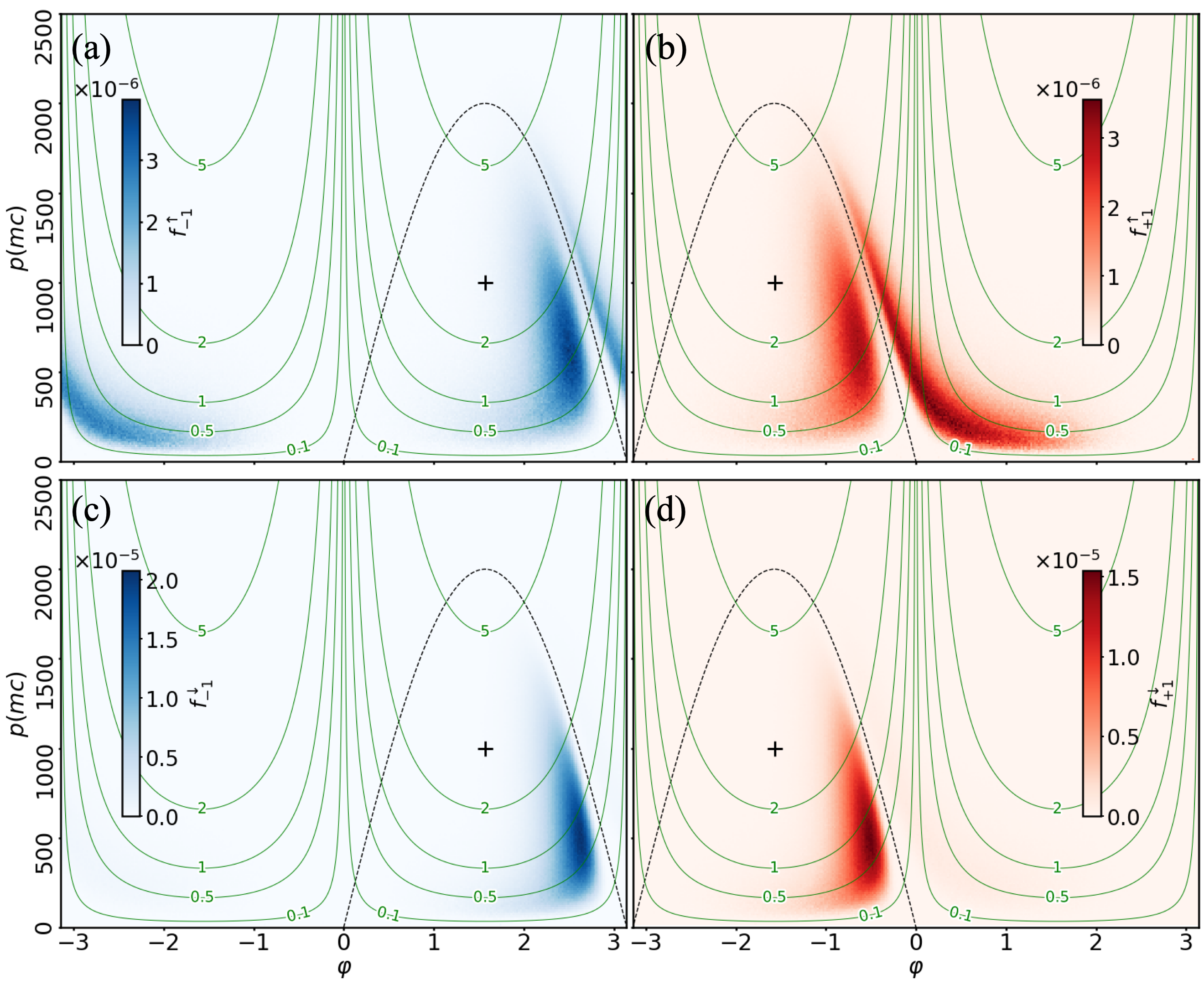}
\centering
\caption{Snapshot of the electron (a), (c) and positron (b), (d) distribution functions in an up (a), (b) or down (c), (d) spin state for the same conditions as the paper calculated using Osiris spin and polarization-dependent QED module.}
\label{fig:PIC_phase_space}
\end{figure}

We started by reproducing the lepton-seeded cascade simulations. Following the conditions given in the paper with laser parameter $a_0 = 1000$ and $\omega = 1.55\ eV$, we obtained the electron and positron distributions shown in Fig.~\ref{fig:PIC_phase_space} using our PIC code. Comparing the result in the paper shown in Fig.~\ref{fig:phase_space}, we can see that our PIC code calculation agrees with the calculation using the Boltzmann-type kinetic equations. For both Figs.~\ref{fig:phase_space} and~\ref{fig:PIC_phase_space}, the spin-down distributions for the electrons and positrons have the highest peak value inside the black dashed line separatrix. This separatrix is the classical advection for the leptons inside the rotation field without radiation energy loss. The spin-related distribution inside the separatrix is dominant by the spin and polarization-resolved NLC process, while the distribution outside the separatrix is dominant by the spin and polarization-resolved NBW pair production process. Due to the difference in the spin up-to-down and down-to-up transition rate of the quantum radiation process, spin-down leptons have a larger population and accumulate inside the separatrix. The particles outside the separatrix come from the pair production process initiated by photons generated from the oppositely charged particles due to their distribution being in different locations in phase space. The pair production process in this simulation generates a similar number of spin-up and spin-down leptons. The spin-up distribution inside the separatrix has a much lower peak than the spin-down distribution, so the distribution outside the separatrix for spin-up leptons is more significant. 


\begin{figure}[h]
\includegraphics[width=8.5cm]{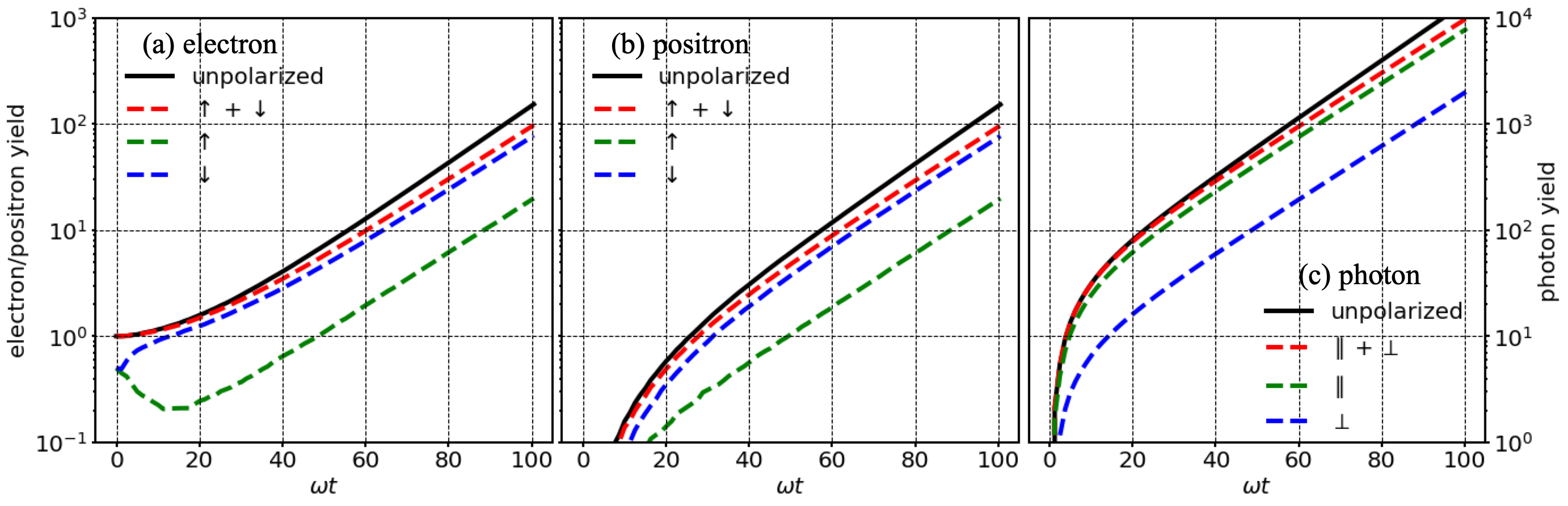}
\centering
\caption{Time evolution of electron (a), a positron (b), and photon (c) yields during a cascade seeded by unpolarized electrons calculated by the Osiris spin and polarization QED module.}
\label{yield_lepton_seeded}
\end{figure}

Fig.~\ref{yield_lepton_seeded} shows the time evolution of electrons, positrons, and photons yield during a cascade seeded with unpolarized electrons calculated using our QED PIC code. Compared with the result in the paper calculated using the Boltzmann-type solver (see figure 3 in Ref.~\onlinecite{Seipt_NJP_2021}), our PIC code gave a similar result. During the cascade process, the quantum radiation process decides the spin distribution of leptons and polarization distribution of photons, while the pair production process decides the growth rate of the leptons. Initially, we can see the number of spin-up leptons decreases. This decrease came from the asymmetry of the spin-flip transition rate in the quantum radiation process. As the cascade process developed, the spin-up to down and down to up transitions were balanced, and the spin-up and spin-down lepton reached a similar growth rate. There is a factor of five times more spin-down leptons than spin-up leptons. The ratio between the photon in different polarization state also become a constant in the exponential growth phase. A factor of four more $\parallel$-polarized photons is emitted compared to $\perp$-polarized photons. Thus, the particles produced in this QED cascade are highly polarized. 

\begin{figure}[h]
\includegraphics[width=7.5cm]{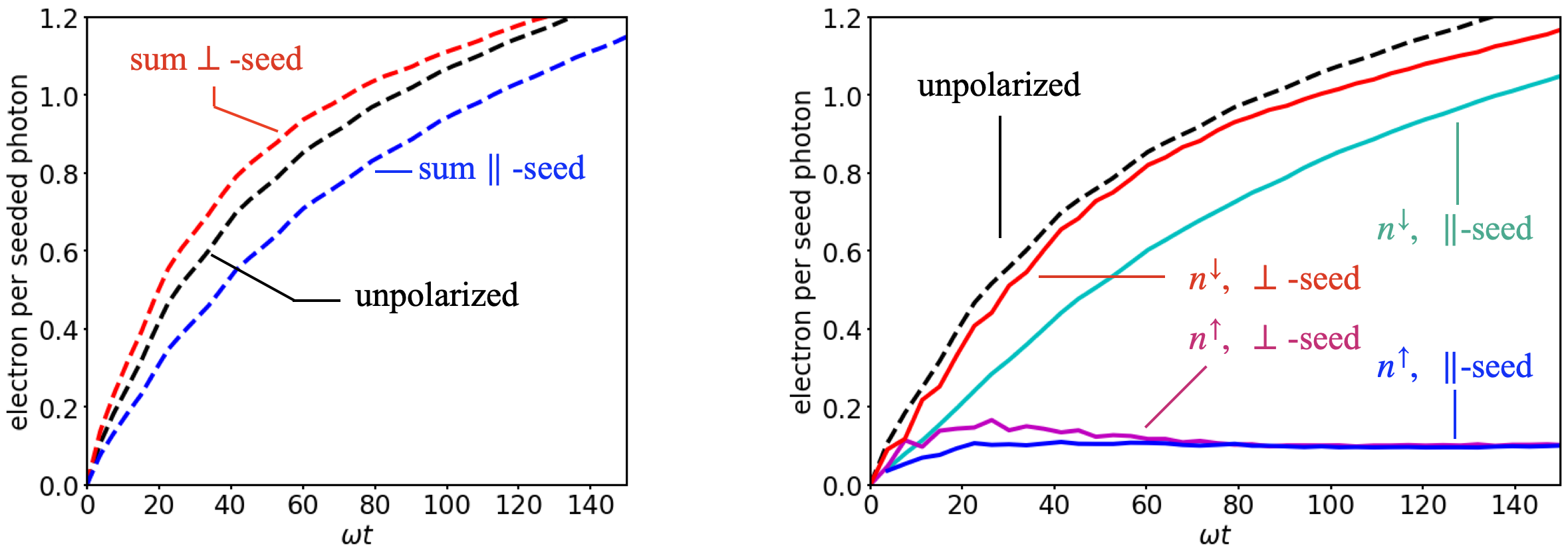}
\centering
\caption{ Time evolution of the electron yield of photon seeded cascade calculated using Osiris spin and polarization QED module.}
\label{yield_photon_seeded}
\end{figure}

Finally, we use our code to reproduce the time evolution of the particle yields for QED-cascade seeded with the polarized photon. Following the same initial conditions in the paper, we calculate the time evolution of the electron yield using our code. The result shown in Fig.~\ref{yield_photon_seeded}  is similar to the result in the paper (figure 5 in Ref.~\onlinecite{Seipt_NJP_2021}). In the left plot, we find that at the early stage, $\perp$-polarized photon-seeded cascade has almost two times high yield of produced pairs than $\parallel$-polarized photon-seeded cascade. As the cascade process developed, the seeding photons were depleted, and the gamma-ray radiated by the lepton dominated the pair production process. The plot on the right shows the yield of spin-down and spin-up electrons. We can see that the $\perp$-polarized photon-seeded cascade generates more spin-down electrons than the $\parallel$-polarized photon-seeded cascade. As time developed, the quantum radiation effect on the lepton spin distribution became dominant, and the number of the spin-down and spin-up particles for both $\perp$-polarized photon-seeded cascade and $\parallel$-polarized photon-seeded cascade would finally become the same.

This section shows that our QED PIC code successfully reproduces the result in the ``Polarized QED Cascades'' paper for both electrons-seeded and polarized photon-seeded cascade simulations. There could be slight differences between our code calculation and the result in the paper. For example, in Figs.~\ref{fig:phase_space} and~\ref{fig:PIC_phase_space}, the color scales for each subplot are different. This difference could come from the intrinsic statistical uncertainty of the Monte Carlo algorithm. Simulating with more particles initially could mitigate this issue.

\section*{Reference}
\bibliography{qed_spin.bib}

\end{document}